\documentclass[
    ,final            
  ]
  {aipproc}

\layoutstyle{8x11double}

\usepackage{amsmath}
\usepackage{graphicx}
\usepackage{dcolumn}
\usepackage{bm}
\usepackage{enumerate}
\usepackage{ctable}
\usepackage{booktabs}
\usepackage{natbib}
\usepackage{siunitx}

\begin{document}

\title{Engaging Physics Faculty in Course Transformation}

\classification{}
\keywords      {change model, three dimensional learning, faculty discussions, shared vision, STEM}

\author{James T.~Laverty}{
  	address={CREATE for STEM Institute, 
  	Michigan State University, East Lansing, MI 48824, USA}
}

\author{Stuart H.~Tessmer}{
  	address={Department of Physics and Astronomy, 
  	Michigan State University, East Lansing, MI 48824, USA}
}

\author{Melanie M.~Cooper}{
	address = {Department of Chemistry,
	Michigan State University, East Lansing, MI 48824, USA},
	altaddress = {CREATE for STEM Institute, 
	Michigan State University, East Lansing, MI 48824, USA}
}

\author{Marcos D.~Caballero}{
	address = {Department of Physics and Astronomy,
	Michigan State University, East Lansing, MI 48824, USA},
	altaddress = {CREATE for STEM Institute, 
	Michigan State University, East Lansing, MI 48824, USA}
}

\begin{abstract}
Research has shown that it is challenging to change departmental culture to support the adoption of transformed teaching practices.
Michigan State University is working to transform its introductory physics courses by engaging faculty in discussions about the scientific practices and core ideas of the discipline, and to use those discussions to drive change to both assessments and classroom instruction.  Here we discuss our model of change, its implementation in the Physics and Astronomy department, the challenges encountered along the way and how we've mitigated those challenges, and tools to measure the impact of this change.

\end{abstract}

\pacs{01.40.G-, 01.40.Gb, 01.55.+b}

\maketitle


\section{Introduction}

Over the past 40 years, physics education research (PER) has developed numerous research-based instructional strategies (RBIS) that improve student learning \cite{Meltzer:2012eg}. However, developing strategies to promote broad usage of RBIS in classrooms has been more difficult \cite{DBERreport, national2013Adapting}.  While most faculty are aware of RBIS, not all of them try to enact them in their courses and those who do often return to traditional teaching practices, usually due to a lack of recognition and/or support \cite{Henderson:2012kl}.

A variety of models to support faculty use of RBIS have been enacted, but most examples are under-documented and their outcomes have been difficult to measure \cite{Dancy:2008up}. In a review of the literature on change, \citeauthor{Henderson:2011ab} noted that an enacted model (a change strategy) in which faculty develop a shared vision for change is most likely to succeed \cite{Henderson:2011ab}. 

The Department of Physics and Astronomy at Michigan State University (MSU) is participating in an interdisciplinary project to transform teaching and learning in introductory STEM courses, funded by the Association of American Universities' STEM Education Intiative. These transformations have been facilitated by a PER postdoctoral fellow, with additional support by two physics faculty, one of whom also conducts PER. Our change strategy uses a shared vision approach where disciplinary faculty attempt to answer two questions: (1) What do we want students to know? (2) What do we want students to be able to do with that knowledge?


As part of this work, we are documenting the process in order to provide generalizable knowledge about change strategies using the work in our department as a typical case. In this paper, we ask: How does our model facilitate the development of a shared vision for not only `how' physics is taught, but also `what' physics is taught? What challenges have developed along the way and how have they been mitigated? How might we objectively evaluate the change process over time?



\section{Implementing a Change Model}

The transformations currently occurring in physics are part of a larger institutional effort that includes chemistry and biology.  The overarching project goal is to transform the culture of the physics, chemistry, and biology departments to value teaching as a scholarly endeavor at the same level as traditional research.


To achieve this goal, we developed a four-pronged approach that is grounded in the literature on institutional change \cite{Henderson:2011ab}.  The four prongs include (1) developing a shared vision within each discipline, (2) using that vision to transform introductory STEM courses, (3) creating a program that recognizes and builds support for excellent teaching, and (4) developing a network of faculty who discuss and share ideas about teaching and learning in STEM. The last two elements are facilitated at the college level and, thus, are outside the scope of the current paper.

\section{Developing a shared vision}

We aim to help the physics faculty develop a shared vision for not only `how' our introductory physics courses should be taught (pedagogy), but also `what' should be taught (content). Our model for change has three principal anchors: (1) articulating `what' should be taught, (2) developing new assessments (e.g., exams and homework) aligned with `what' should be taught, and (3) changing `how' courses are taught to align with `what' should be taught. Not only is the knowledge students should have at the end of the course open for discussion, but so is how that knowledge should be put into practice.

\begin{figure}[t]
\includegraphics[width=\columnwidth]{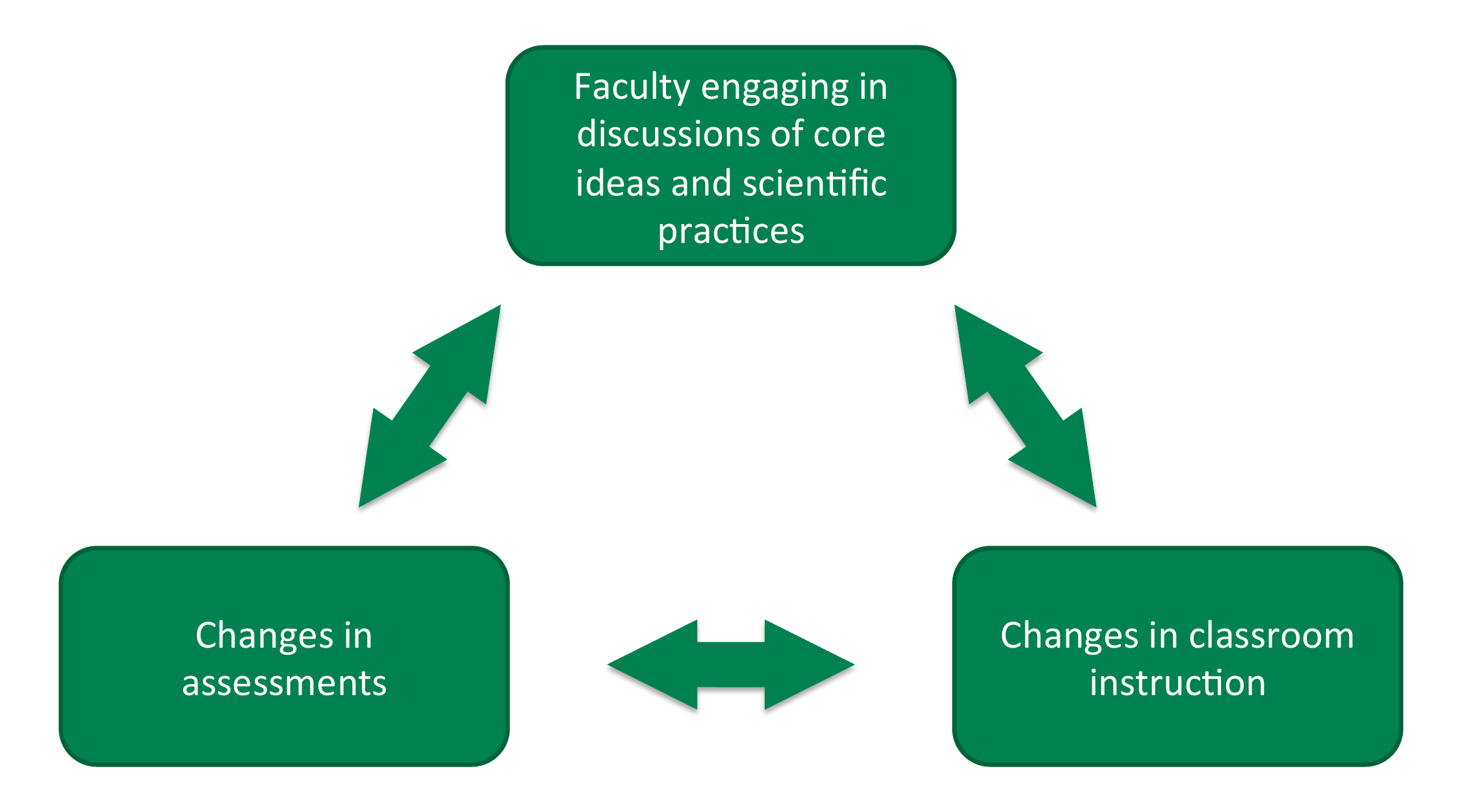}
\caption{Our model for change uses faculty discussions about core ideas and scientific practices to drive changes to assessments and instruction.}
\label{fig:changeassessinstruct}
\end{figure}

To articulate `what' should be taught, our model engages faculty in discussions to answer two questions: (1) What do we want students to know? and (2) What do we want students to be able to do with that knowledge? We leverage these discussions to determine what faculty value and the answers to these questions are developed into goals (performance expectations). These performance expectations are reflected in the development of new assessments, which are designed to assess what our faculty value. Changing the assessments is critical because doing so will impact how the introductory courses are taught (i.e., pedagogy) as faculty align their instruction with the new goals. The linear progression described here is not completely accurate, as each anchor of the model influences the other as shown in Fig.~\ref{fig:changeassessinstruct}.

\subsection{Framework for Change}

The framework for the faculty discussions employs the idea of three dimensional learning (3DL), which is at the core of the Next Generation Science Standards \cite{Standards:2012ur, NGSS}.  Our faculty discussions bring 3DL to the university level as a means to develop and to modify the performance expectations, assessments, and instruction in the introductory physics courses. The dimensions of 3DL emphasize not only the body of knowledge that is the result of science, but also the process of science \cite{Standards:2012ur}:


\begin{description}
\item[Scientific Practices] Knowledge of how science achieves its ends. The ability to engage with the scientific endeavor. {\it Example: Developing and using models}
\item[Crosscutting Concepts] Concepts that bridge disciplinary boundaries. These ideas appear in each of the disciplines. {\it Example: Systems and system models}
\item[Disciplinary Core Ideas] Concepts that are essential to the study of a discipline, are required to explain a wide range of phenomena, and provide a way to generate new ideas and predictions. {\it Example: Matter and its interactions}
\end{description}

A key point of 3DL is that no dimension should be taught in isolation from the others.  The goal is to blend all three dimensions into all teaching and learning opportunities, including lectures, homework, in-class activities, and exams.

\begin{figure}[t]
\includegraphics[width=\columnwidth]{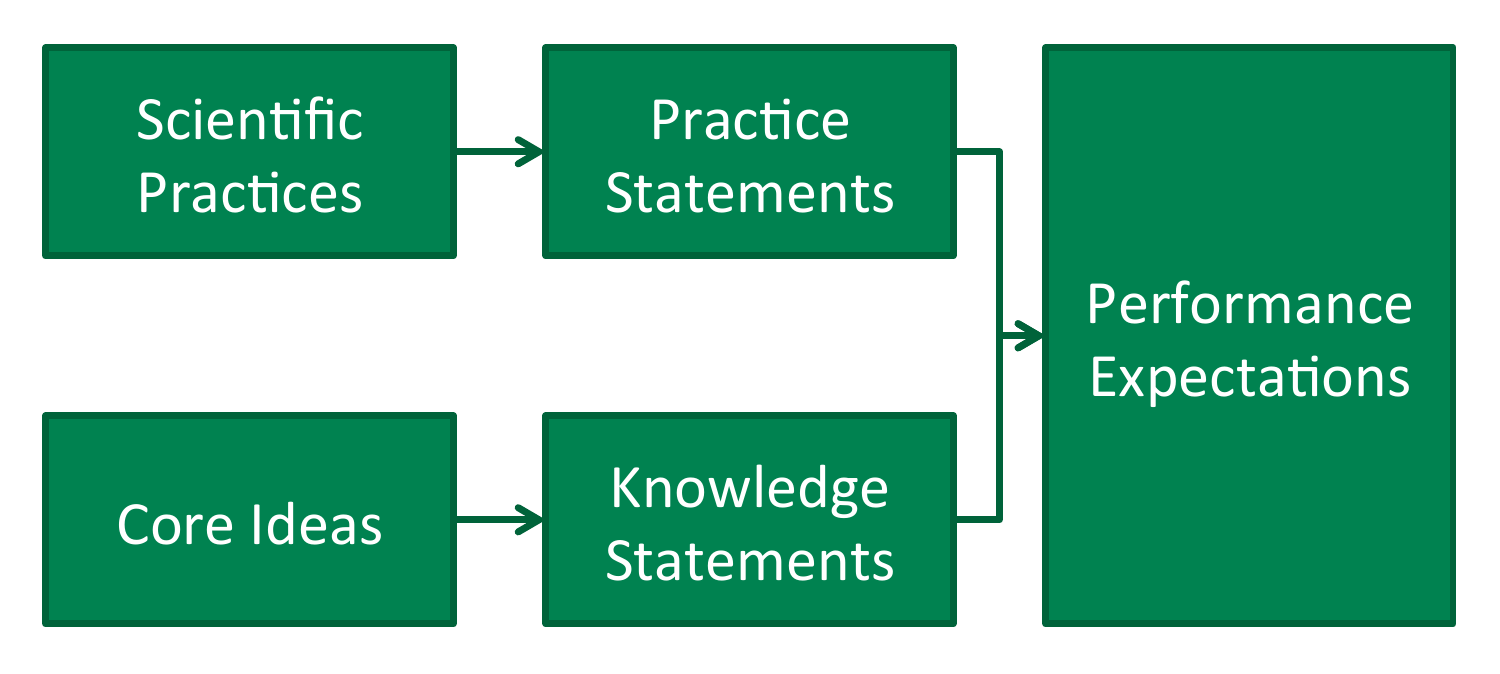}
\caption{Framework for conducting faculty discussions.  The process begins by identifying the scientific practices and core ideas in the introductory courses and then proceeds to blend these into a set of performance expectations for the course.}
\label{fig:discussionframework}
\end{figure}

The framework for our discussions within the physics department focus on scientific practices and physics core ideas. Eventually, groups with faculty members from physics, chemistry, and biology will discuss crosscutting concepts and how they might be implemented across the introductory science courses (if they aren't already). To date, the physics faculty have started blending the practices and core ideas they have established into performance expectations (Fig.~\ref{fig:discussionframework}) from which the new assessments will be developed.



\section{Facilitating a shared vision}

In the Physics and Astronomy department, a postdoctoral research associate is facilitating the faculty discussions and helping the faculty develop performance expectations for the introductory courses using the process shown in Fig.~\ref{fig:discussionframework}. These performance expectations are statements that describe what students should be able to do with their physics knowledge when they finish the course. Assessments aligned with these expectations will be developed with the faculty and used as metrics for student learning.




Early in the process, it is useful for faculty to construct a ``complete'' list of scientific practices and core ideas that they believe are important in each course.  This process provides a negotiated list, which the faculty can build on through the development of ``practice statements" and ``knowledge statements''. These ``statements'' are the steps in which the faculty articulate a more detailed description of the scientific practices and core ideas.  For instance, our faculty established the core idea of {\it Energy, Heat, \& Work} for introductory mechanics and developed a number of knowledge statements related to that core idea (e.g., {\it Energy of a system only changes due to work done on/by the system and heat exchanged with the surroundings}).  The faculty have also developed several practice statements for each of the scientific practices, such as {\it Engage in Evidence-based Arguments} (e.g., {\it Argue for the efficacy of one experimental design over another to test a hypothesis}). The next step for the discussions is to define the performance expectations by blending these items together.

We have used Fig.~\ref{fig:discussionframework} to anchor our faculty discussions and to move toward the development of performance expectations that blend scientific practices and core ideas.  This framework has not only clarified the goals of the faculty discussions, but also helped to focus individual discussions on specific elements.  For instance, while the primary goal is to determine the performance expectations, the agenda for one discussion was to try to define the practice statements associated with {\it Constructing and Using Models}.

\section{Lessons Learned}

Throughout this work, we have encountered a number of challenges that could have disengaged faculty from the process. These challenges were mitigated by the facilitator, the faculty, or both and, thus, we have kept faculty engaged. Below, we discuss a number of lessons that we have learned along the way that may be helpful in guiding other institutions in the future.  A broad theme of these lessons is that the input and perspectives of the faculty are important and should be respected.


\textbf{Explain the process early and often.} The process described in this paper is designed to re-envision the introductory courses from the ground up.  A common reaction among the faculty early in the process was to pick specific elements of a course that were relatively easy to modify (e.g., choosing a new textbook or adding a few clicker questions).  While these choices are important, we reminded faculty that the goal was to design the department's course as a whole. Concerns about the textbook, homework problems, demonstrations, et cetera could not be productively discussed until the goals of the course were made explicit.  In doing this, we were careful not to shut the faculty down, but to defer such conversations to a later time when they would be more productive.


\textbf{Listen to faculty.} The faculty must decide what is important in these classes.  We have focused on keeping the discussion productive within the confines of the process.  If the faculty felt the facilitator was ignoring their ideas or that they were not being heard, they would lose interest in the process.  Faculty input has been used to modify the process (see below) as well as to determine the agenda and nature of individual discussions. 

\textbf{Allow faculty to modify the process.} The initial discussions to determine the core ideas and scientific practices were open to any and all faculty that were interested.  After these first two discussions, members of the faculty and administration expressed an interest in the discussions becoming more official.  Three subcommittees were formed to formalize the process. Because of the size of the department (\textasciitilde65 faculty), the standard practice for making decisions about the undergraduate program is to use committees. These subcommittees are comprised of interested faculty and faculty who often teach these courses with membership spanning the various professorial levels (instructor, assistant, associate, full). While formalizing this process limits the faculty who can engage in the process, it respects a cultural norm of the department that legitimizes the process in ways a more informal discussion could not.


\textbf{...but not too much.} Certain elements of the process can be negotiated because they are influenced by situational factors, but other aspects are central to the model (Fig.~\ref{fig:discussionframework}) and should not be changed.  We found it important to discuss with faculty why these aspects are critical without resorting to statements similar to, ``because that's what research has shown,'' which can cause faculty to disengage.  For instance, on several occasions faculty suggested that core ideas should be taught in the lecture courses and scientific practices should be taught in the lab courses.  We have consistently reminded faculty that blending these dimensions results in students being able to use their knowledge in ways the faculty appreciate.

\textbf{Provide examples.} We found it helpful to provide clear examples of the various aspects of the process (Fig.~\ref{fig:discussionframework}).  Early on, the faculty struggled to define performance expectations using the scientific practices and core ideas they had developed.  After that discussion, the facilitator blended one of the scientific practices and one of the core ideas into a sample performance expectation, which was discussed the next time the faculty met.  Having a concrete example moved the process along and improved the quality of the conversations substantially. At the close of the discussion, a faculty member volunteered to develop another example. Providing faculty with examples has helped jump start the process and has been crucial to the faculty engaging in productive conversations.

\textbf{Introduce new language slowly.} We have found that new language that faculty might perceive as jargon can pose a barrier to faculty engagement. A productive approach has been introducing new terms as a short-hand way of describing something (which is usually how such terms are created).  Early in the process, we avoided the term ``scientific practices'', which our faculty perceived as ``jargony.'' Instead, we asked our faculty, ``What would you like students to be able to do with their knowledge?''  After a few discussion meetings, this language became too cumbersome and we suggested shortening it to simply ``practices.'' Since this introduction, faculty have taken up this language during the discussions.

\textbf{Support change from the outside.} At MSU, grants, support from administration, and outside speakers have helped move the process along.  Our faculty are interested in teaching well, but many are not informed about the improvements seen at peer institutions or how to achieve such improvements. To help educate our faculty about what other departments have done, we have held two physics colloquia in the past year given by PER faculty from other institutions.  Both speakers were well received and their talks have helped generate conversations with additional faculty about what steps the department might take to improve.

\section{Assessing the impact of the shared vision}

In addition to documenting the change process and the lessons that might be abstracted for other future efforts, we are working with other discipline-based education researchers to assess the impact of our change process.
This effort focuses on the changes resulting in the courses by analyzing how the products of the model (the in-class instruction and the assessments; see Fig.~\ref{fig:changeassessinstruct}) align with 3DL.

To investigate changes in instruction, each section of the introductory courses has been recorded three times during the semester over the past year.  This has provided \textasciitilde 30 hours of recordings in physics alone.  In order to analyze this data, we are developing the Three Dimensional Learning Observation Protocol (3D-LOP).  This tool looks at both the `how' (employed pedagogy) and `what' (inclusion of 3DL) of instruction. We are in the process of validating the protocol.

In order to determine how in-class assessments have changed, the exams from each section of every introductory course have been collected and will continue to be collected for several years.  In parallel to the 3D-LOP, we are also developing the Three Dimensional Learning Assessment Protocol (3D-LAP), which analyzes assessment items such as exams, homework, and clicker questions to determine if they contain scientific practices, crosscutting concepts, and disciplinary core ideas. The coding structure for this tool is being finalized.

Our plan is to use both the 3D-LAP and 3D-LOP to investigate how the assessments and classroom instruction change over the lifetime of the project and in the years beyond.

\section{Conclusion}

This paper has focused on the implementation of a model that has the goal of changing departmental culture to value the scholarly activity of teaching.  As part of this model, we are engaging faculty in the Department of Physics and Astronomy at MSU in discussions about the scientific practices and core ideas that are important in introductory physics.  These discussions are currently in the process of developing performance expectations for those courses and will soon be working to align the assessments and instruction in those courses with the newly defined performance expectations.  Several lessons were presented that can be used to guide other institutions in the future and instruments for assessing the goals of the change were briefly introduced.\\



The authors thank the Association of American Universities' STEM Education Initiative for their support, the faculty members of the Physics and Astronomy Department at MSU for their continuing effort to improve student learning, and the rest of the DBER community at MSU, especially those involved in these efforts and PERL@MSU for their feedback on early drafts.

\bibliographystyle{aipproc}   
\bibliography{thehitlist}

\end{document}